\documentclass[journal=jctcce,manuscript=article,floatfix]{achemso}

\setkeys{acs}{doi=true}
\usepackage{graphicx}
\usepackage{dcolumn}
\usepackage{bm}
\usepackage{subcaption} 

\usepackage[T1]{fontenc} 
\usepackage{hyperref} 
\hypersetup{
    colorlinks=true,
    linkcolor=red,
    filecolor=magenta,      
    urlcolor=blue,
    citecolor=blue,
    }
\newcommand{\doi}[1]{\href{http://dx.doi.org/#1}{\nolinkurl{#1}}}
\usepackage{amsmath}
\usepackage{amssymb}

\title{Machine learning classification of local environments in molecular crystals}

\author{Daisuke Kuroshima}
\affiliation{Department of Chemistry, New York University (NYU), New York, New York 10003, USA.}
\email{daisuke.kuroshima@nyu.edu}

\author{Michael Kilgour}
\affiliation{Department of Chemistry, New York University (NYU), New York, New York 10003, USA.}
\email{michael.kilgour@nyu.edu}

\author{Mark E. Tuckerman}
\affiliation{Department of Chemistry, New York University (NYU), New York, New York 10003, USA.}
\alsoaffiliation{Courant Institute of Mathematical Sciences, New York University, New York, New York 10012, USA.}
\alsoaffiliation{NYU-ECNU Center for Computational Chemistry at NYU Shanghai, 3663 Zhongshan Rd. North, Shanghai 200062, China.}
\alsoaffiliation{Simons Center for Computational Physical Chemistry at New York University, New York, New York 10003, USA.}
\email{mark.tuckerman@nyu.edu}

\author{Jutta Rogal}
\affiliation{Department of Chemistry, New York University (NYU), New York, New York 10003, USA.}
\alsoaffiliation{Fachbereich Physik, Freie Universit{\"a}t Berlin, 14195 Berlin, Germany.}
\email{jutta.rogal@nyu.edu}

\date{\today}
\begin{document}             
\begin{abstract}
Identifying local structural motifs and packing patterns of molecular solids is a challenging task for both simulation and experiment. 
We demonstrate two novel approaches to  characterize local environments in different polymorphs of molecular crystals using learning models that employ either flexibly learned or handcrafted molecular representations. 
In the first case, we follow our earlier work on graph learning in molecular crystals, deploying an atomistic graph convolutional network, combined with molecule-wise aggregation, to enable per-molecule environmental classification.
For the second model, we develop a new set of descriptors based on symmetry functions combined with a point-vector representation of the molecules, encoding information about the positions as well as relative orientations of the molecule.
We demonstrate very high classification accuracy for both approaches on urea and nicotinamide crystal polymorphs, and practical applications to the analysis of dynamical trajectory data  for  nanocrystals and solid-solid interfaces. 
Both architectures are applicable to a wide range of molecules and diverse topologies, providing an essential step in the exploration of complex condensed matter phenomena.
\end{abstract}

\maketitle

\section{\label{sec:introduction}Introduction}

Elucidation of the microscopic structure of molecular materials is key to predicting and engineering their properties. 
Despite significant advances in experimental techniques, following structural transformations in condensed-phase systems with atomistic resolution remains a challenge due to the time- and length-scales involved. 
Computational approaches, such as molecular dynamics (MD) simulations, have become an invaluable tool to provide such microscopic insight, but characterizing 
the structural features of a molecular system from the simulation data is, in general, nontrivial.
However, following the dynamical evolution of local structural environments is essential when studying polymorphic transitions, especially regarding the complex atomistic processes that govern nucleation and growth.

A number of descriptors have been developed over the years to capture local or global structural features, including Steinhardt order parameters~\cite{steinhardt1983bond,lechner2008accurate}, common 
neighbor analysis~\cite{honeycutt_MolecularDynamics_1987,faken1994systematic,stukowski2012structure}, entropy based fingerprints~\cite{piaggi2017entropy}, smooth overlap of atomic positions~\cite{bartok2013representing}, and atom-centred symmetry functions~\cite{behler2007generalized} (see also~\cite{neha2022collective,tanaka2019revealing,tong2014order,yang2016structures,reinhardt2012local,eslami2017local,piaggi2017enhancing,song2020generating} for further overviews and  examples). 
More recently, machine learning has been utilized to classify local environments with both supervised and unsupervised approaches~\cite{geiger2013neural,cubuk2015identifying,rogal2019neural,rogal2021pathways,defever_GeneralizedDeep_2019,boattini2018neural,boattini2020autonomously,scheiber2022binary,bapst2020unveiling,banik2023cegann,ishiai2023graph,ishiai_GraphNeuralNetworkBasedUnsupervised_2024}.
These machine learning models for local structure classification fall into two broad categories: models that use handcrafted structural features or descriptors together with a simple classification model, and models that use only very general information, such as atom types and distances, and letting the model learn the structural representation and intermolecular correlations simultaneously.
The former approach is attractive in its ostensible simplicity but relies on the development of high-quality descriptors, while the latter requires a more complex model architecture but less intuition about the system and is more generally applicable.
Here, graph neural-network (GNN) approaches are attractive in their generality, allowing one to use a single flexible model for most systems.
GNNs have also been used to describe condensed-phase systems, wherein the relevant features are learned in a `ground up' fashion from basic atomistic information \cite{xie2018crystal,park2020developing,kim2020gcicenet,ishiai2023graph,banik2023cegann,beyerle2023recent,chen2019graph,zhou2020graph,wu2020comprehensive, fung2021benchmarking}.

The structure characterization methods discussed above have primarily been established in the context of atomistic condensed-matter systems. 
In molecular systems, additional challenges arise since not only the positions of the molecules but also their relative orientation as well as conformational changes need to be accounted for.
One idea is to include this information via a point-vector representation of the molecules where, for example, the center of mass denotes the molecule position, and vectors denote the absolute orientation of a given molecule, which can then be combined into suitable descriptors~\cite{santiso2011general,shah2011computer}.

In this work, we advance the state of the art of machine learning classification of local environments to capture the complex structural features in molecular solids. We present two parallel approaches, one based on handcrafted descriptors and the other on learned feature embeddings. 
The handcrafted descriptors extend our previous work on atomistic systems~\cite{rogal2019neural} to molecular symmetry functions (SF) by combining the SFs  with a point-vector representation of the molecules. For the learned embeddings, we utilize our recently introduced molecular crystal graph model MXtalNet~\cite{kilgour2023mxtalnet} and augment the architecture with a classification task.
The trained models are able to distinguish different local environments in various polymorphs of complex molecular solids with high accuracy. 
Furthermore, both approaches are applicable to a wide range of systems, including clusters and interfaces, and can provide time-resolved information regarding melting transitions or solid-solid transformations.
The potential of our classification models is exemplified for urea and nicotinamide but the methods are easily extended to arbitrary molecules. The approaches presented introduce an essential and valuable component in the analysis and interpretation of simulation data for molecular solids.

\section{\label{sec:methods}Model architectures and training}

\begin{figure*}
    \includegraphics[width=\textwidth]{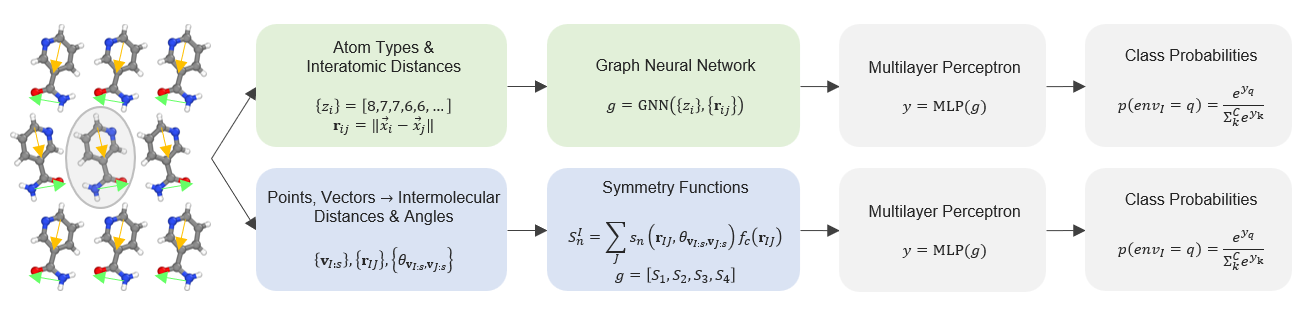}
    \caption{\label{fig:scheme} The workflow of the GNN and SF classifiers on top and bottom, respectively, including molecule representation, local embedding, and classification. The GNN learns the features $g$ used in the classification task, while for the SF classifier the features $g$ are given by the handcrafted molecular SFs.} 
\end{figure*} 
The general idea of our two model architectures is schematically illustrated in Fig.~\ref{fig:scheme}. The classification is performed for each molecule to characterize its local structural environment. 
An appropriate model should be invariant to permutations of atoms of the same types, as well as global translations, rotations, and inversions of the atomic coordinates, focusing only on the structural correlations which define the respective polymorphs. 
For the learned feature embedding, the positions and atom types of a given molecule and its neighbors comprise the input to a graph neural network (GNN) coupled with a multilayer perceptron (MLP) to perform classification on the final embedding. 
For the handcrafted features, the atomic positions are used to construct a point-vector representation for each molecule which is then employed to compute a set of molecular symmetry functions as input to the classification MLP. 
Details of the model architectures and training protocols are given in the following. 

\subsection{Molecular crystal graph network} 
For the molecular GNN, we used a relatively straightforward graph neural network, similar in geometric complexity to SchNet~\cite{schutt2018schnet}, taking interatomic distances and atom types as inputs.
This GNN encodes pairwise interatomic distances to edge embeddings, atom types to node embeddings, and performs graph convolutions via the TransformerConv operator~\cite{shi2020masked} implemented in the Torch Geometric package~\cite{Fey/Lenssen/2019}.

The GNN parses a single sample in the following way, starting with embedding of the input nodes atom types $z_i$,
\begin{equation}\label{eq:G1}
    \mathbf{f}_i^0 = \mathrm{EMB}(z_i) \quad,
\end{equation}
with EMB as a learnable discrete embedding function, followed by a fully-connected layer.
The edge embedding is
\begin{equation}\label{eq:G3}
    \mathbf{e}_{ij} = \mathrm{Bessel}(|\mathbf{r}_{ij}|) \quad,
\end{equation}
where $|\mathbf{r}_{ij}|$ is the distance between nodes $i$ and $j$, and Bessel is the radial embedding function from DimeNet~\cite{gasteiger2020directional} with cutoff $r_c=6$~\AA\ and 32 radial Bessel basis functions.
A fully connected layer is defined as $\text{FC}(\mathbf{x})=\mathbf{W}\cdot \mathbf{x} + \mathbf{b}$, with $\mathbf{W}$ and $\mathbf{b}$ as learnable parameters. 
Messages are passed between nodes, conditioned on node and edge embeddings via Eqs.~\eqref{eq:G4} for edge$\to$message and \eqref{eq:G5} for node$\to$message, over $N$ graph convolutions, with GC the graph convolution operation,
\begin{equation}\label{eq:G4}
    \mathbf{E}^t_{ij}=\mathbf{W}^t_{e\rightarrow m}\cdot\mathbf{e}_{ij} \quad,
\end{equation}
\begin{equation}\label{eq:G5}
    \mathbf{F}_i^t=\mathbf{W}^t_{n\to m}\cdot\mathbf{f}_i^t \quad,
\end{equation}
\begin{equation}\label{eq:G6}
    \mathbf{f}^{t+1}_i=\mathbf{W}^t_{m\to n}\cdot(\mathrm{GC}(\mathbf{F}_i^t,\mathbf{F}_j^t,\mathbf{E}_{ij}^t)) \quad.
\end{equation}
After each graph convolution, the node embeddings are passed through a fully-connected layer with residual connection,
\begin{equation}\label{eq:G6.5}
    \mathbf{f}^{t}_i=\mathbf{f}^{t}_i+\sigma\left(D\left(\mathcal{N}(\mathrm{FC}^t_{n\to n}(\mathbf{f}^{t}_i))\right)\right) \quad,
\end{equation}
with $\sigma$ being the activation function (here GeLU~\cite{hendrycks2016gaussian}), $D(x)$, a dropout function, and $\mathcal{N}(x)$, the graph layer norm operation. 
The final node features, corresponding to information about each atom and its local environment, are aggregated into a single embedding vector representing the entire molecule, and input to a two-layer activated fully-connected network with layer normalization and dropout, followed by a reshaping to the  number of possible classes. 
Though there are currently many powerful graph aggregators, we find max aggregation, that is, selecting the maximum value from each feature channel, $k$, across the final atomic node embeddings in each molecule, is simple and efficient for learning the desired functions, with
\begin{equation}\label{eq:G7}
    \mathbf{g}=\mathrm{MAX}_k(\{\mathbf{f}^N\})
\end{equation}
and
\begin{equation}\label{eq:G8}
    \mathbf{y}=\mathrm{MLP}(\mathbf{g}) \quad,
\end{equation}
with $\mathrm{MLP}$ a multilayer perceptron. 
The class probabilities for a molecule $I$ being in a particular environment $q$ are computed via the softmax activation function
\begin{equation}\label{eq:G9}
    p(\text{env}_I = q)=\frac{\exp(y_q)}{\sum_k^{C} \exp(y_k)} \quad,
\end{equation}
with $C$ the number of possible environments.

We found one or two graph convolutions gave similar performance, though more convolutions result in a larger volume for what the model considers as a `local environment'.
The number of convolutions depends on the user's desired sensitivity to longer-range structural correlations, but in the current examples,
more than two convolutions resulted in training instability and overall poor convergence.
For other hyperparameters, optimal performance was obtained with a relatively deep embedding (256 for node and graph embeddings, 128 for messages), aggressively regularized with layer norm and dropout of 0.25 in graph convolutions, nodewise fully-connected layers, and the embedding-to-output network. 
With these settings, the model  converged via the Adam optimizer to a the test minimum very quickly, usually within a few tens of epochs. 
Smaller models could certainly be explored, although we generally found convergence properties to be poorer in that regime.
For further details of model construction, see the supplementary information (SI) and our accompanying codebase~\cite{mkgnngit}.

\subsection{Molecular symmetry functions}
\begin{figure}
\centering
\includegraphics[width=0.5\textwidth]{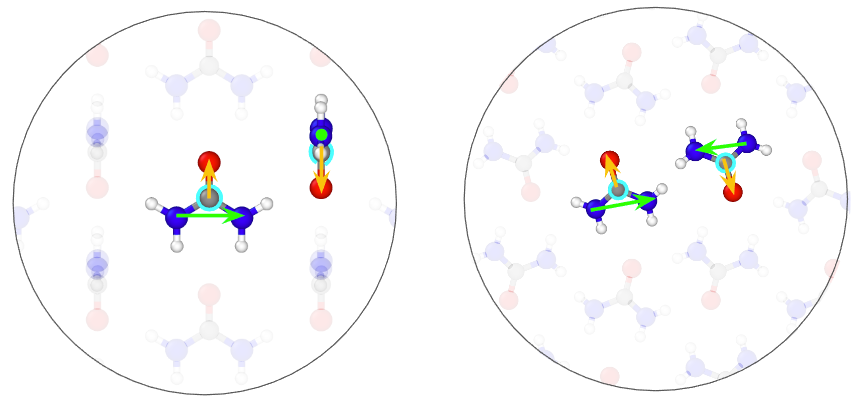} \\[2ex]
\includegraphics[width=0.5\textwidth]{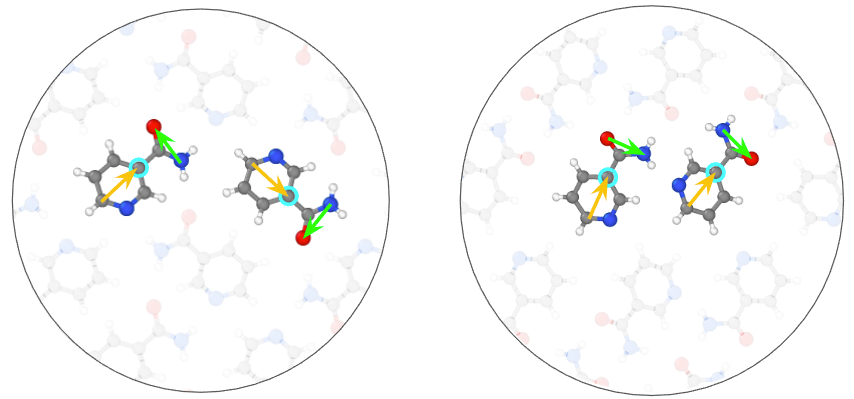}
\caption{\label{fig:ureaarrows} Point-vector representation for urea (top panels)  and nicotinamide (bottom panels) in two different polymorphs, respectively. The turquoise circles indicate the positions of the molecules $\mathbf{r}_I$, and the green and orange vectors, $\mathbf{v}_{I;1}$ and $\mathbf{v}_{I;2}$, characterize their relative orientations.}
\end{figure}
Our second model derives a set of descriptors for each molecule based on the Behler-Parrinello symmetry functions~\cite{behler2007generalized} in combination with a point-vector representation~\cite{santiso2011general,shah2011computer} of the molecules.
The point-vector representations for urea and nicotinamide are illustrated in Fig.~\ref{fig:ureaarrows}, where the position $\mathbf{r}_I$ of molecule $I$ is represented by a selected atom (indicated by a turquoise circle in Fig.~\ref{fig:ureaarrows}). Vectors $\mathbf{v}_{I;s}$ are defined between two selected atoms in the molecule, such that they can capture relative orientations of the molecules (indicated in orange, $\mathbf{v}_{I;1}$, and green, $\mathbf{v}_{I;2}$, in Fig.~\ref{fig:ureaarrows}).
We utilize four different types of molecular symmetry functions $S^I$. Two are akin to radial symmetry functions for atomistic systems but using the molecule positions $\mathbf{r}_I$, 
\begin{equation}\label{eq:S1}
    S_{1}^I(\mathbf{r}) = \sum_{J}{e^{-\eta \ (|\mathbf{r}_{IJ}| - R_{s})^2}f_{c}(\mathbf{r}_{IJ})} \quad,
\end{equation}
and
\begin{equation}\label{eq:S2}
    S_{2}^I(\mathbf{r}) = \sum_{J}{\cos(\kappa {|\mathbf{r}_{IJ}|)f_{c}(\mathbf{r}_{IJ})}} \quad,
\end{equation}
where the sum runs over all other molecules, $\mathbf{r}_{IJ} = \mathbf{r}_{J} - \mathbf{r}_{I}$, $f_c$ is a cutoff function (see SI for details), and $\eta$, $R_s$, and $\kappa$ are tunable parameters.
The other two types of molecular symmetry functions use the molecule vectors to characterize the relative orientation of molecule $I$ with respect to its neighbours $J$, 
\begin{equation}\label{eq:S3}
    S_{3}^I(\mathbf{r},\mathbf{v}_{;s}) = \sum_{J}{\exp\left(-\eta \ (\cos\theta_{\mathbf{v}_{I;s}\mathbf{v}_{J;s}} - \cos\theta_{S})^2\right)f_{c}(\mathbf{r}_{IJ})} \quad,
\end{equation}
and
\begin{equation}\label{eq:S4}
     S_{4}^I(\mathbf{r},\mathbf{v}_{;s}) = \sum_{J}{\cos\left(\kappa \cos\theta_{\mathbf{v}_{I;s}\mathbf{v}_{J;s}}\right)f_{c}(\mathbf{r}_{IJ})},
\end{equation}
where $\theta_{\mathbf{v}_{I;s}\mathbf{v}_{J;s}}$ is the angle between vectors $\mathbf{v}_{;s}$ on molecules $I$ and $J$, and $\cos\theta_{S}$ is another tunable parameter. 
The total number of molecular symmetry functions is 24 for both urea and nicotinamide. 
Details regarding the selected molecular symmetry functions and corresponding values of the tunable parameters are given in the SI.

To perform classification of molecule environments, the molecular symmetry function descriptors are input to a rather small MLP with two hidden layers, 25 nodes each, and the softmax activation in Eq.~\eqref{eq:G9} for the output layer. 
A larger MLP with more hidden layers and nodes would provide greater flexibility but 
due to the simplicity of the classification task,  a small network was sufficient for our applications, making both the training and evaluation rather fast. 
For further implementation details, see the SF classifier codebase~\cite{dksfgit}.

\subsection{Training the models}
Training data were generated by molecular dynamics (MD) simulations of all crystal polymorphs and the melt for urea and nicotinamide.
Simulations were performed using the \textsc{lammps} MD package~\cite{LAMMPS} with a general Amber force field (GAFF)~\cite{wang2004development}.
We briefly summarize here the protocol for training the classification models. 
Further details regarding the MD simulations and training are given in the SI.

The graph classifier was trained on a mix of trajectory snapshots of periodic bulk cells approximately $20\times20\times20$~\AA$^3$ and  gas phase spherical clusters with a diameter of  $\sim 30$~\AA\ 
to give the effect of a `surface'.
Molecules are identified as being on the surface if their local coordination number, $\text{CN}_{I}$, is smaller than 20, with $\text{CN}_{I} = \sum_{J} \theta(-(|\mathbf{r}_{IJ}| - R_C))$, where $\theta$ is the Heaviside function and $R_c$ the molecule radius plus the graph convolution cutoff.
The symmetry function classifier was trained on periodic bulk samples only.

We train the classification models on stable, low-temperature snapshots of the known polymorphs of each molecular crystal, as well as a higher temperature melt state.
We test the models' generalization performance on configurations from  higher temperature MD simulations, with adaptation to thermal noise standing as a proxy for overall generalization.
The specific temperatures for each of the studied systems are discussed together with the results below.

The graph classifier was trained until the test loss began to increase, and the model checkpoint at test loss minimum was used for evaluation.
Repeated retraining over several random seeds found variation in test loss minimum of only a few percent between runs. 
We used a combined cross entropy loss including both the loss for the local polymorph classification for each molecule and the molecule topology, that is `surface' vs `bulk'.

The symmetry function classifier was trained until the training loss converged which, generally, resulted in very small test losses.

\section{\label{sec:results}Classification of local environments}

\subsection{Bulk polymorphs of urea and nicotinamide}
\begin{figure}
    \includegraphics[width=0.7\textwidth]{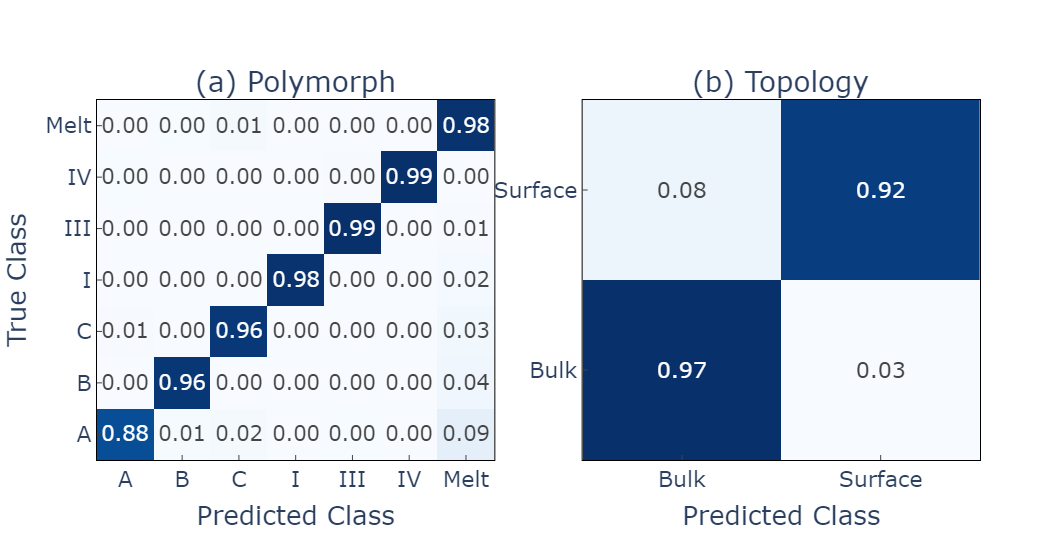}
    \caption{\label{fig:urea_accuracy} Confusion matrix for the graph classifier on (a) the polymorphs  and (b) topologies of urea at 200K for crystals and 350K for melt. Micro F1 scores=0.969, 0.960.}
\end{figure} 
We initially trained and applied our classification models to two different systems, urea and nicotinamide. 
Urea is a relatively small and rigid molecule, which is also significantly polymorphic, having six distinct crystal structures with unique intermolecular packing character~\cite{swaminathan1984crystal, olejniczak2009h,giberti2015insight, shang2017crystal} (see the SI for a visualization of the respective polymorphs).
The models were trained on $T=100$~K crystal samples and $T=350$~K~melts, and evaluation metrics were computed on  samples at 200~K for the crystal polymorphs and 350~K for the melt.
At low temperatures, the graph classifier achieves perfect accuracy for both polymorphs and local topologies.
This means that the GNN learns an embedding where the different molecule environments are clearly separated without overlap. 
This is expected as the graph model is rather expressive and in all the thousands of individual molecular environments, the local structure seen by the model should be quite similar within each polymorph.
The graph model also generalizes well to higher temperature samples at $T=200$~K, as evidenced by the confusion matrix shown in Fig.~\ref{fig:urea_accuracy}, meaning that larger thermal fluctuations can be captured within the trained model.
Only urea form A shows a slighter larger classification error, with about 9\% of the samples being identified as `melt', which might be due to the lower stability of form A.
The symmetry function classifier also demonstrated excellent performance on urea, achieving comparable or better performance at polymorph classification ($F1\gtrsim 0.98$) to the GNN model in training and evaluation while being lightweight and fast to run at inference. The corresponding confusion matrix can be found in the SI.

\begin{figure}
    \includegraphics[width=0.7\textwidth]{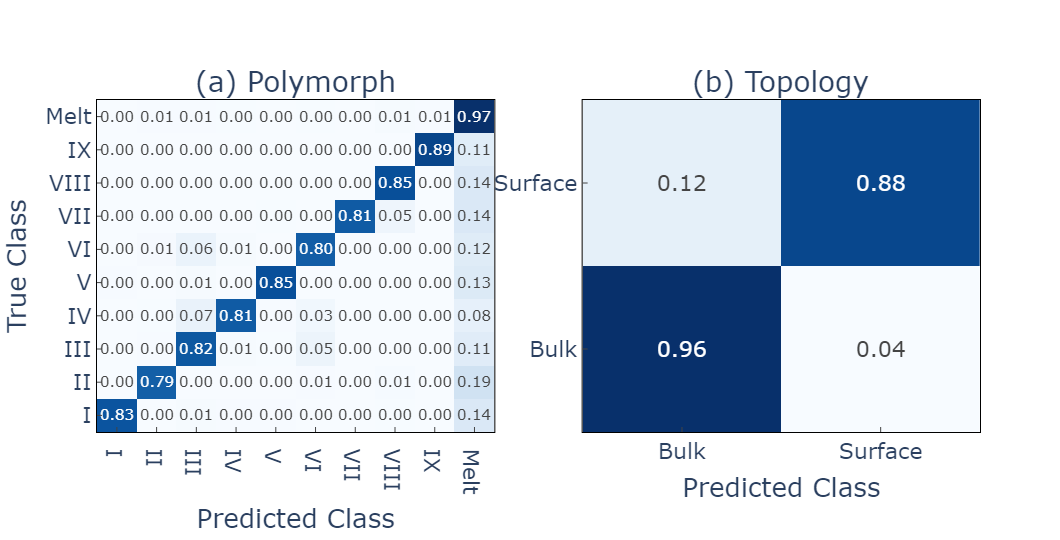}
    \caption{\label{fig:nic_accuracy} Confusion matrix for the graph classifier on (a) the polymorphs and (b) topologies  of nicotinamide at 350~K. Micro F1 scores=0.875, 0.922.}
\end{figure} 
As a second example, we chose nicotinamide as a more challenging molecule.
Nicotinamide is larger than urea and more flexible with internal degrees of freedom that allow for polymorphs consisting of different conformers of the molecule. Nine polymorphs of nicotinamide have been experimentally crystallized~\cite{li2020rich,fellah2021highly}(see the SI for a visualization of the respective polymorphs).
Despite this significant added complexity in the molecular system, the performance of our classification models is again very good.
As with urea, the training samples, both crystal polymorphs at 100~K and melts at 350~K, are essentially perfectly learned, and the model generalizes well to the high temperature test samples at 350~K.
The corresponding confusion matrix for the GNN classifier is shown in Fig.~\ref{fig:nic_accuracy}.
The F1 score for nicotinamide at high temperatures is slightly worse than for urea, 0.875 compared to 0.969, which reflects the increased flexibility in the thermal fluctuations at this even higher temperature.
This is, however, not a fundamental limit of the model, as, when retrained across the full range of temperatures, the accuracy again approaches 100\%.

We see that the generality and high capacity of the GNN model allow it to classify each polymorph and local topology, without the need for model customization of any kind.
Likewise, the symmetry function classifier performs excellently on the nicotinamide polymorphs (see the SI for the corresponding confusion matrices). This indicates that the chosen set of molecular symmetry function provides suitable descriptors to capture the additional complexity and flexibility in nicotinamide crystal polymorphs and melt.

One interesting point is that the GNN classifier exhibits a somewhat lower performance on the nicotinamide high temperature samples compared to the SF classifier, when both are trained on low temperature crystals and high temperature melts only. 
From the confusion matrix in Fig.~\ref{fig:nic_accuracy} it becomes clear that the accuracy loss of the graph classifier is primarily due to  over-prediction of the melt state. 
For a model trained only at 100~K and evaluated at 350~K, this should perhaps not be surprising. 
The larger thermal fluctuations in inter- and intramolecular degrees of freedom increase the general similarity of bulk crystals to the melt, and they are interpreted as such by the model.
That we do not see this effect as strongly in the SF classifier results indicates that the handcrafted descriptors are quite robust to fluctuations, yet sensitive enough to achieve high classification accuracy.

\begin{figure*}
    \includegraphics[width=\textwidth]{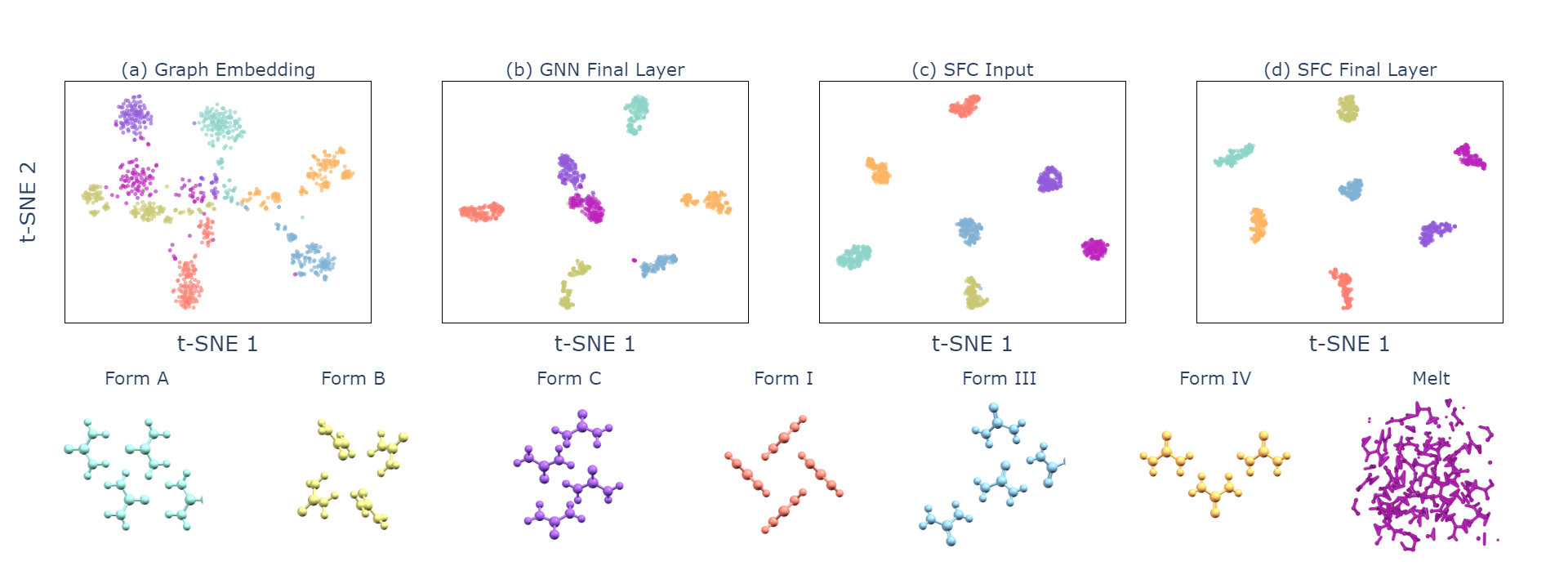}
    \caption{\label{fig:urea_tsne} The t-distributed stochastic neighbor embedding (t-SNE) of urea samples from (a) the 256-dimensional graph embedding (output of \eqref{eq:G7}), (b) 256-dimensional final layer activation, (c) 24 symmetry functions, and (d) 25-dimensional SFC final layer activation; samples are taken from three different temperatures of 100~K, 200~K, and 350~K.} 
\end{figure*}  
To get a better understanding of the learned and handcrafted features in our molecular graph and symmetry function classifiers, respectively, we compare the corresponding embedding spaces.
In Fig.~\ref{fig:urea_tsne}, the embedding spaces of the representations and final layer activations for urea are visualized using the t-distributed stochastic neighbor embedding (t-SNE)~\cite{hinton2002stochastic}.
Fig.\ref{fig:urea_tsne}(a) shows that the molecular representation learned by the GNN already separates the different polymorphs of urea reasonably well. 
The quality of the handcrafted symmetry functions is obvious when examining the t-SNE of the symmetry function inputs in Fig.~\ref{fig:urea_tsne}(c), which cluster essentially perfectly before applying any learned transformations.
Figs.\ref{fig:urea_tsne}(b) and~(d) show the t-SNE of the final layer activations for the GNN and symmetry function classifier, respectively.  The class separation is excellent, as  expected from the very high classification accuracy observed for both models.

\begin{figure*}
    \includegraphics[width=\textwidth]{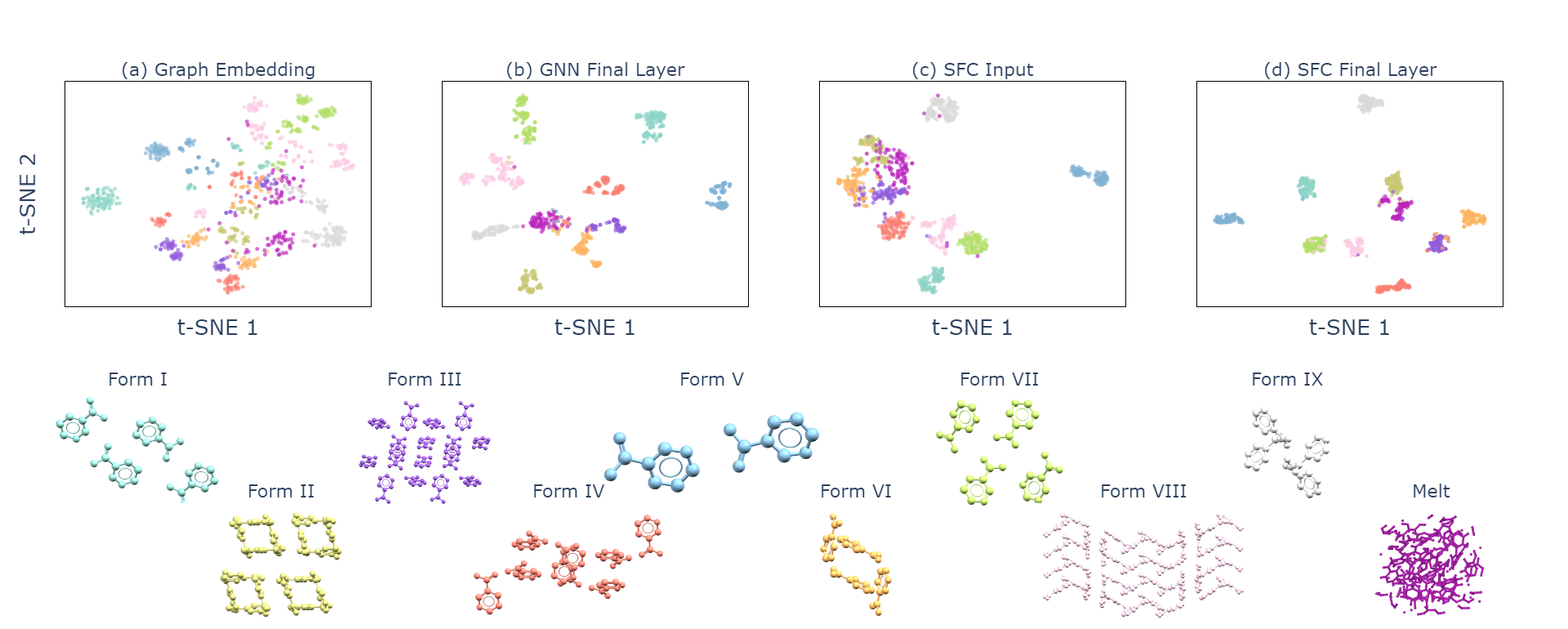}
    \caption{\label{fig:nic_tsne} t-SNE of nicotinamide samples from (a) the graph embedding (output of \ref{eq:G7}), (b) final layer activation, (c) symmetry functions, and (d)  SFC final layer activation, at temperatures of 100~K and 350~K. Embedding dimensionality is the same as in Figure \ref{fig:urea_tsne}.} 
\end{figure*}
The t-SNE visualization of the embedding spaces for nicotinamide are shown in Fig.~\ref{fig:nic_tsne}.
Both the learned and handcrafted embedding spaces in Figs.~\ref{fig:nic_tsne}(a) and~(c) show imperfect classwise separation between the various polymorphs in nicotinamide. 
This again underscores the increased challenge in characterizing structural environments in more complex and flexible systems. In particular, samples from the melt seem to cover a wide range and are less clustered in the embedding spaces.
We also see greater separation of samples from the same crystalline polymorphs in Figs.~\ref{fig:nic_tsne}(a)-(b), including bifurcation of some classes, corresponding to the different sampled temperatures and topologies.
The overlap between the melt and crystal embeddings visible in Figs.~\ref{fig:nic_tsne}(a)-(b) is also consistent with the GNN classifier confusing some crystalline polymorphs mainly with the melt, as seen in Fig.~\ref{fig:nic_accuracy}.
Nevertheless, the final learned representations in Figs.~\ref{fig:nic_tsne}(b) and~(d) show again a very good separation between the different polymorph classes, even for the high-temperature samples.

\subsection{Analyzing molecular simulations}
Being able to reliably characterize local environments in unknown structures will be particularly useful when analyzing and interpreting trajectory data from molecular simulations. 
In the following, we discuss two examples: the stability of gas phase nanocrystals at different temperatures and the migration of an interface during a solid-solid transformation in a molecular crystal. 

\subsubsection{Dynamical structure characterization of molecular clusters}
\begin{figure}
    \includegraphics[width=0.8\textwidth]{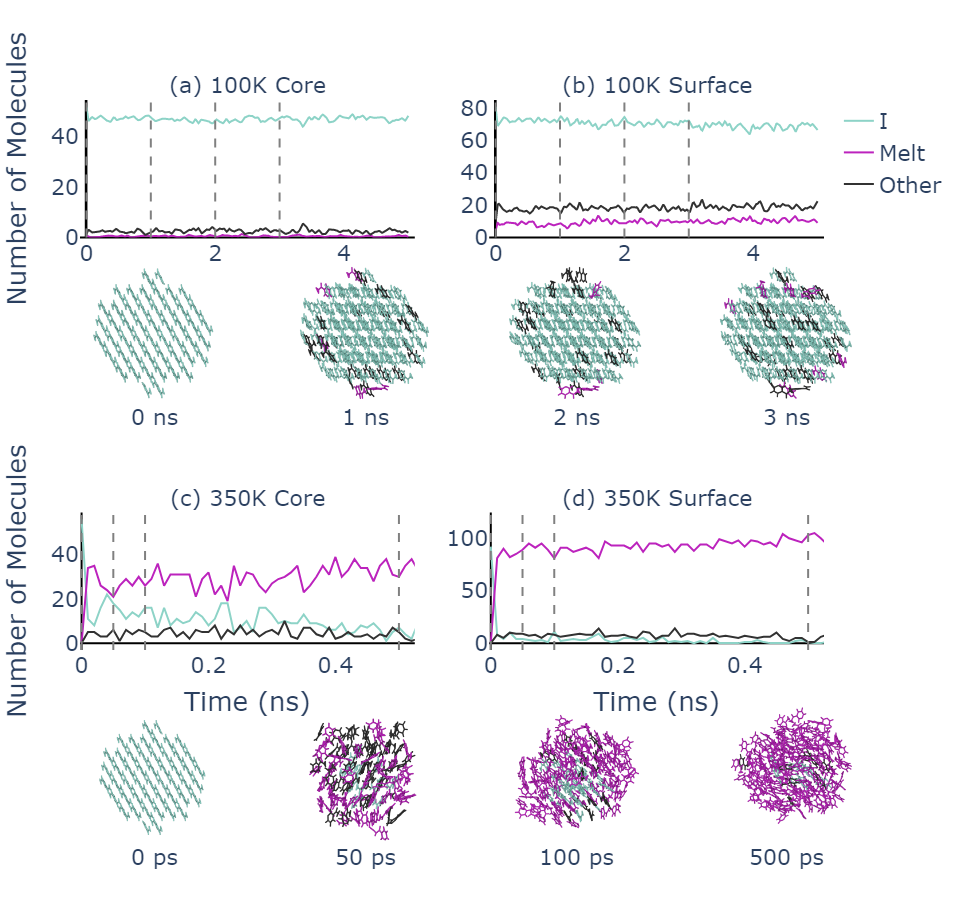}
    \caption{\label{fig:nic_trajs} Time evolution of the number of molecules classified as form I, melt, or other, (a)-(b) at 100~K and (c)-(d) at 350~K. The analysis is shown separately for high-coordination `core' molecules in (a) and (c) and low-coordination `surface' molecules in (b) and (d). 
    Vertical dashed lines identify the time points for the cluster snapshots, with molecules coloured according to their most probable form. Snapshots were visualized using \textsc{ovito}~\cite{stukowski2009visualization}
    }
\end{figure}

The GNN classifier trained on the bulk polymorphs of nicotinamide is used to identify the local environments of nicotinamide molecules in small nanocrystals. 
We set up spherical clusters of nicotinamide form I with a diameter of 34~\AA\ containing 148 molecules. 
Molecular dynamics simulations for the clusters in vacuum are run at $T=100$~K and 350~K (further simulation details are given in the SI).
In Fig.~\ref{fig:nic_trajs}, the structural evolution of the nicotinamide nanoclusters at these two temperatures is shown, obtained using the graph classifier.
Since the classifier provides information for each molecule individually, we can separate our analysis for molecules that are in the core region of the clusters, Fig.~\ref{fig:nic_trajs}(a) and~(c), and at their surfaces,  Fig.~\ref{fig:nic_trajs}(b) and~(d).
At 100~K, the nanocluster clearly keeps its crystalline structure over the entire simulation time. While the majority of molecules in the core region are identified as nicotinamide form I, molecules at the surface are partially classified as melt or others, which is expected since the structural environment at the surface is significantly different from the bulk.
At 350~K, the crystalline cluster quickly melts starting from the surface. Within a few picoseconds, molecules at the surface are identified as liquid with a handful labeled as others. 
The core region melts a little more slowly with a few molecules initially remaining as form I and others.
After approximately $500$~ps, the cluster appears to be completely melted with only a small number of core molecules identified as others.

Despite not having been trained on clusters in vacuum or mixtures of polymorphs, the performance of our graph classifier in the analysis of the simulation data is sensible and very informative, allowing to evaluate the structural stability and the onset of melting as a function of temperature.

\subsubsection{Time evolution of solid-solid phase boundaries}
\begin{figure}
    \includegraphics[width=0.8\textwidth]{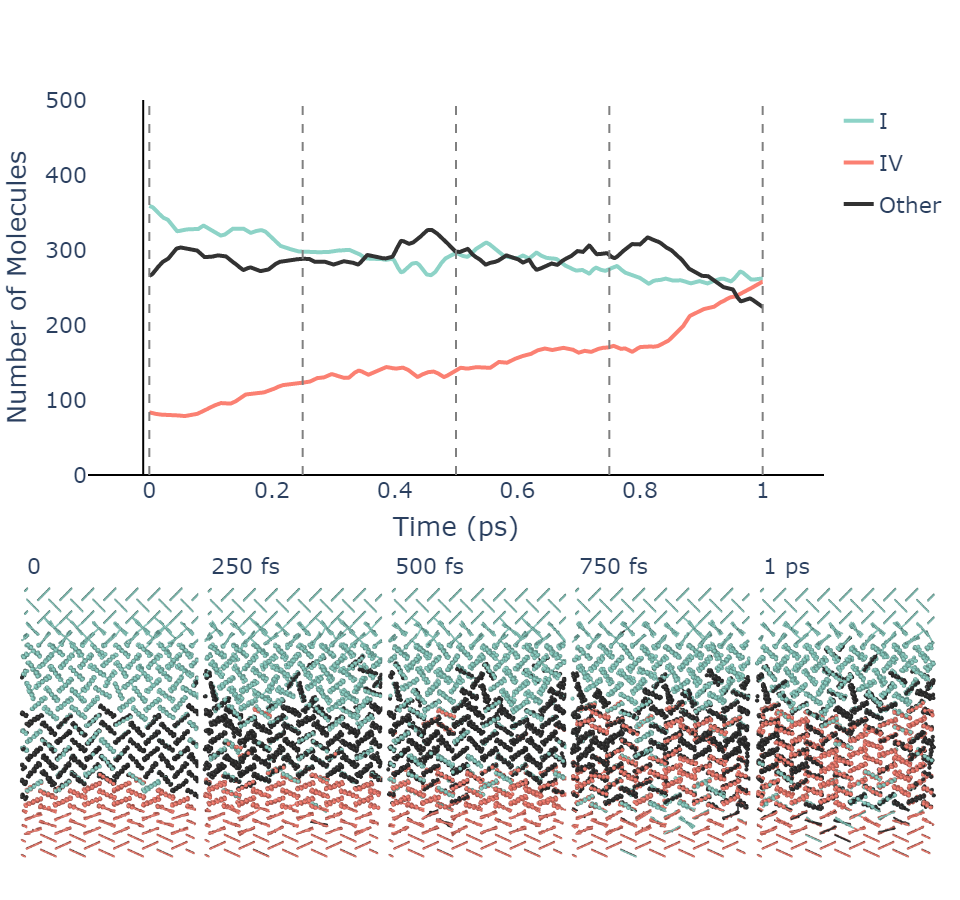}
    \caption{\label{fig:urea_traj} Time series of the molecule-wise composition of a system with a moving interface between form I and IV of urea. In the top graph, only molecules in the central region of the simulation cell, highlighted in bold in the snapshots below, are included. 
    Vertical dashed lines correspond to the time points from which the snapshots were sampled, with molecules coloured according to their assigned polymorph. Snapshots were visualized using \textsc{ovito}~\cite{stukowski2009visualization}.}
\end{figure}
Pushing our analysis tools even further, we apply our classification models to track the position of the interface between two different polymorphs of urea during a solid-solid transformation.
A semi-coherent interface between form I and IV of urea is set up by pairing both phases along the $[001]$  direction. The $xy$-dimensions parallel to the interface are fixed resulting in  1.7\% compression in $x$ and 1.4\% strain in $y$ of urea I and 2.8\% compression in $x$ and 0.8\% strain in $y$ of urea IV, respectively. 
Periodic boundary conditions are applied in all dimensions, keeping molecules at one of the interfaces fixed and simulations are run in the NVAP$_z$ ensemble at $T=100$~K (further simulation details are given in the SI).

In Fig.~\ref{fig:urea_traj}, the analysis of the structural transformation using the graph classifier is shown. 
Initially, the system is mainly composed of urea form I (green molecules) in the top half of the simulation cell with some form IV at the bottom. Molecules at the interface between the two polymorphs are primarily identified as `others' due to deviations of their local environments from the pure bulk polymorphs. 
Since within the chosen setup form I is rather unfavourable, transformation to form IV rapidly takes place over a few hundred femtoseconds, which is indicated by the continued increase of molecules identified as form IV and decrease of form I in the top graph of Fig.~\ref{fig:urea_traj}.

Here again, the utility of accurate local environment classification is clearly evidenced, as subtle changes in local spacing and orientations of molecules can be seen to correspond to the transformation between distinct polymorphs, in this case form I and IV of urea.
Interestingly, we also see that the conversion from form I to IV is not perfect, as some defects are left in the wake of the phase boundary as it moves upward through the sample.

\section{\label{sec:summary}Conclusions}
We have introduced two machine learning based approaches  for the classification of local structural environments in molecular solids.
Both the GNN classifier with  learned feature embeddings and the SF classifier with handcrafted descriptors identify molecular environments in various bulk polymorphs with high accuracy.
While the performance of the two machine learning models is comparable for the studied systems, there are differences in their practical application.

The GNN model can be used for most molecular systems `out of the box' with minimal customization but may require hyperparameter tuning to achieve good generalization.
Due to its flexibility and expressive power, with the model presented here containing 356k parameters, the GNN classifier is somewhat sensitive to overfitting the training data.
Again, one could train a smaller GNN model, at the empirically observed cost of slower convergence to inferior evaluation minima.
Still, the model evaluates relatively quickly, at ~35 training iterations, each comprising some hundreds of molecules, per second on V100 GPU compute and $\sim$1 per second on a single CPU.
During evaluation, the performance bottleneck is more often the conversion from MD trajectory output files into the appropriate data format for the GNN model than the model forward pass itself, with  500 trajectory frames of ~20~\AA$^3$~ bulk systems taking usually only several minutes to analyse.
For sufficiently complex problems, a GNN classifier could in the future be upgraded with more sophisticated geometric features, convolutional methods or global aggregators, to efficiently capture longer-range intra- and inter-molecular dependencies within a particular system.
Today, such architectural improvements are relatively well understood and adoptable `off the shelf'.

The performance of the SF classifier strongly depends on the handcrafted input features. The molecular symmetry functions proposed here do provide the flexibility to capture complex environments in molecular solids but need to be carefully chosen for each new system. This includes both the point-vector representation of the respective molecule as well as the tunable parameters of the symmetry function. For larger and more flexible molecules, it might be necessary to expand the molecular symmetry functions to explicitly account for conformational changes, for example, by introducing symmetry functions that depend on different vectors in the same molecule.
It is, however, desirable to keep the number of molecular SFs small since calculating the input descriptors is the main computational cost when evaluating the SF classifier.

Both models are trivially parallelizable as they only require the information for a given molecule and its environment. They are also applicable to multi-component systems, such as co-crystals, or can be used to identify defects, such as impurities, vacancies, surfaces, or interfaces. The main challenge in these more complex scenarios is the preparation of labeled training data for the supervised learning task.

The two classification models presented in this study provide a general approach for the analysis and interpretation of simulation data in molecular solids. This will be particularly useful for the study of structural transformations, including nucleation and growth.  
Additionally, information about the local environment can also be used to construct collective variables used in enhanced sampling of structural transformations, as we have shown previously for atomistic systems~\cite{rogal2019neural,rogal2021pathways}.
We expect that the characterization of local structural motifs  using classification models will become an essential tool in the simulation of molecular solids, as these models are easy to train and extremely versatile. 

\begin{acknowledgement}
The authors would like to thank Leslie Vogt-Maranto for fruitful discussions.
The work of MK was supported by a Natural Science and Engineering Research Council of Canada (NSERC) post-doctoral fellowship. JR acknowledges financial support from the Deutsche Forschungsgemeinschaft (DFG) through the Heisenberg Programme project 428315600. JR and MET acknowledge funding from grants from the National Science Foundation, DMR-2118890, and MET from CHE-1955381. This work was supported in part through the NYU IT High Performance Computing resources, services, and staff expertise.
\end{acknowledgement}

\begin{suppinfo}
The supplemental information contains details regarding the descriptors, architecture, and training of the machine learning models as well as regarding the molecular dynamics simulations to create training and test data and molecular simulation examples.    
\end{suppinfo}

\bibliography{references}

\end{document}


\title{Supplementary Information \\ Machine learning classification of local environments in molecular crystals}

\author{Daisuke Kuroshima}
\affiliation{Department of Chemistry, New York University (NYU), New York, New York 10003, USA.}

\author{Michael Kilgour}
\affiliation{Department of Chemistry, New York University (NYU), New York, New York 10003, USA.}

\author{Mark E. Tuckerman}
\affiliation{Department of Chemistry, New York University (NYU), New York, New York 10003, USA.}
\affiliation{Courant Institute of Mathematical Sciences, New York University, New York, New York 10012, USA.}
\affiliation{NYU-ECNU Center for Computational Chemistry at NYU Shanghai, 3663 Zhongshan Rd. North, Shanghai 200062, China.}
\affiliation{Simons Center for Computational Physical Chemistry at New York University, New York, New York 10003, USA.}

\author{Jutta Rogal}
\affiliation{Department of Chemistry, New York University (NYU), New York, New York 10003, USA.}
\affiliation{Fachbereich Physik, Freie Universit{\"a}t Berlin, 14195 Berlin, Germany.}

\maketitle

\section{Crystal structures of urea and nicotinamide polymorphs}
\begin{figure}[h]
    \begin{center}
    \includegraphics[width=0.9\textwidth]{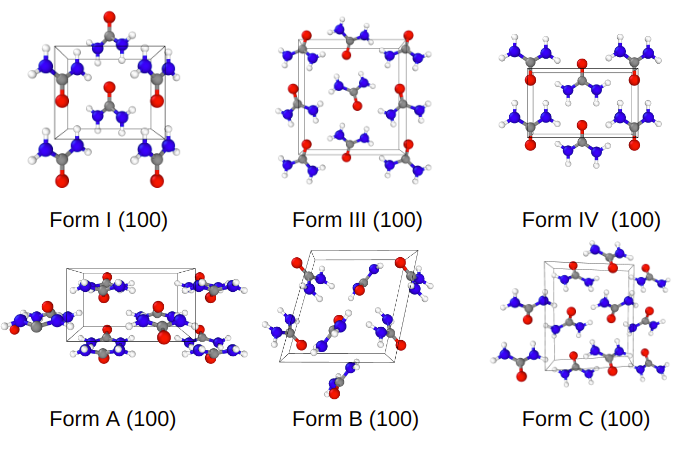}
    \caption{\label{fig:urea_all} Crystal structures of six urea polymorphs used in this study viewed along the [100] direction,
     including the experimentally crystallized forms I, III, and IV, as well as computationally proposed forms A, B, and C.} ~\cite{swaminathan1984crystal,olejniczak2009h,giberti2015insight,shang2017crystal}
    \end{center}
\end{figure}

\begin{figure}[h]
    \begin{center}
    \includegraphics[width=0.9\textwidth]{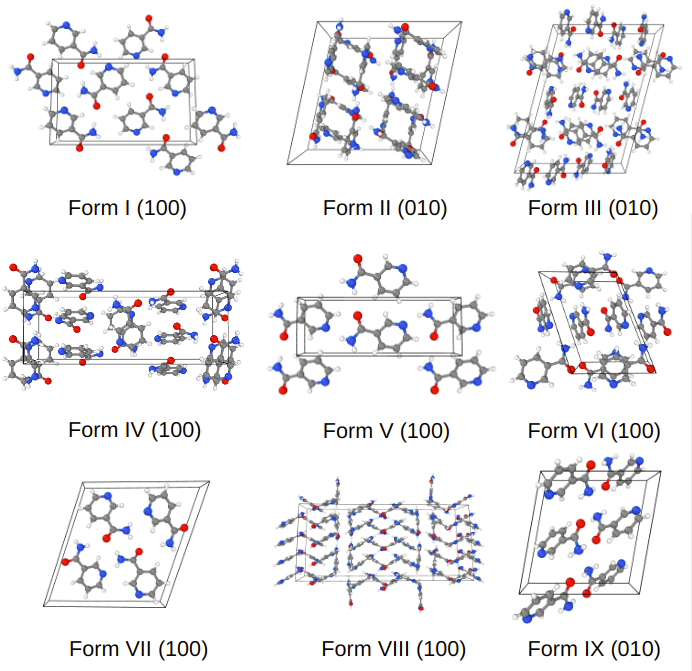}
    \caption{\label{fig:nicotinamide_all} Crystal structure of nine nicotinamide polymorphs used in this study. Form I, IV, V, VII, and VIII are viewed along the [100] direction, 
    and Form II, III, VI, and IX along [010].
    }
    \end{center}
\end{figure}

Figs.~\ref{fig:urea_all} and~\ref{fig:nicotinamide_all} visualize the different polymorphs of urea and nicotinamide, respectively. These structures have been visualized using Ovito~\cite{stukowski2009visualization}.

\clearpage
\section{Molecular symmetry functions and Training}
The cutoff function used in the molecular symmetry functions has the following form~\cite{chen2012heating}
\begin{equation}
  f_c(\mathbf{r}_{IJ})=\begin{cases}
    1 & \text{if} \; |\mathbf{r}_{IJ}| < r_\text{min}\\
    \frac{1}{2} \left(\cos\left[ \frac{(|\mathbf{r}_{IJ}| - r_{min})}{r_{c} - r_{min}} \pi\right]+1\right) & \text{if} \;  r_\text{min}<|\mathbf{r}_{IJ}| \leq r_c \\
    0 & \text{if} \; |\mathbf{r}_{IJ}| > r_c
  \end{cases}
\end{equation}
where $|\mathbf{r}_{IJ}|$ is the distance between  molecule $I$ and $J$. The  cutoff radii are set to $r_\text{min}$ = 9.8~\AA\ and  $r_c$ = 10.0~\AA\ for urea and $r_\text{min}$ = 6.8~\AA\  and  $r_c$ = 7.0~\AA\ 
 for nicotinamide. A set of input function was carefully selected by computing the distributions of  symmetry function values for a series of the tunable parameters $R_s$, $\cos\theta_S$, $\eta$, and $\kappa$. The overlap of distributions for different polymorphs were compared  and parameters resulting in small overlaps  were selected. In total, 24 molecular symmetry functions  
 were selected for both urea and nicotinamide. The corresponding values for the parameters are given in Tab.~\ref{Table:urea} for urea and Tab.~\ref{Table:nicotinamide} for nicotinamide.

\begin{table}[ht]
\centering
\caption{\label{Table:urea}Parameters of the molecular symmetry functions used for urea. 
}
\begin{tabular}{ |p{3cm}||p{3cm}|p{3cm}|p{3cm}|p{3cm}|p{3cm}|}
 \hline
 symmetry function& $R_s$&$\cos\theta_S$ &$\eta$&$\kappa$&vector\\
 \hline
 $S_{1}^I$&&&&& \\
 1&6.16&-&2.44&-&- \\
 2&6.28&-&2.68&-&- \\
 3&6.76&-&1.00&-&- \\
 4&6.88&-&1.00&-&- \\
 \hline
 $S_{2}^I$&&&&&\\
 5&-&-&-&2.50&- \\
 6&-&-&-&4.54&- \\
 7&-&-&-&4.90&- \\
 8&-&-&-&6.22&- \\
 \hline
 $S_{3}^I$&&&&&\\
 9&-&0.368&1.00&-&C-O \\
 10&-&0.08&1.00&-&C-O \\
 11&-&0.36&1.12&-&C-O \\
 12&-&0.28&6.76&-&C-O \\

 13&-&-0.64&3.28&-&N-N \\
 14&-&-0.36&3.28&-&N-N \\
 15&-&0.88&3.28&-&N-N \\
 16&-&1.00&3.28&-&N-N \\
 
  \hline
 $S_{4}^I$&&&&&\\
 
17&-&-&-&2.50&C-O \\
 18&-&-&-&3.58&C-O \\
 19&-&-&-&4.78&C-O \\
 20&-&-&-&8.26&C-O \\
 21&-&-&-&2.50&N-N \\
 22&-&-&-&8.12&N-N \\
 23&-&-&-&8.24&N-N \\
 24&-&-&-&8.36&N-N \\
 \hline
\end{tabular}
\end{table}%

\begin{table}[ht]
\centering
\caption{\label{Table:nicotinamide}Parameters of the molecular symmetry functions used for nicotinamide. 
}
\begin{tabular}{ |p{3cm}||p{3cm}|p{3cm}|p{3cm}|p{3cm}|p{3cm}|}
 \hline
 symmetry function &$R_s$&$\cos\theta_S$ &$\eta$&$\kappa$&vector\\
 \hline
 $S_{1}^I$&&&&& \\
 1&3.75&-&1.26&-&- \\
 2&5.25&-&0.01&-&- \\
 3&4.9&-&0.1&-&- \\
 4&5.9&-&0.016&-&- \\
 \hline
 $S_{2}^I$&&&&&\\
 5&-&-&-&1.06&- \\
 6&-&-&-&0.51&- \\
 7&-&-&-&1.03&- \\
 8&-&-&-&2.41&- \\
  \hline
 $S_{3}^I$&&&&&\\
 9&-&0.01&0.66&-&C-C \\
 10&-&1.66&0.01&-&C-C \\
 11&-&2.9&0.9&-&C-C \\
 12&-&1.69&4.0&-&C-C \\

 13&-&0.56&3.0&-&O-N \\
 14&-&0.66&0.01&-&O-N \\
 15&-&1.18&17.2&-&O-N \\
 16&-&1.15&2.2&-&O-N \\
 
 \hline
 $S_{4}^I$&&&&&\\
 17&-&-&-&0.13&C-C \\
 18&-&-&-&0.33&C-C \\
 19&-&-&-&5.05&C-C \\
 20&-&-&-&3.2&C-C \\
 21&-&-&-&1.05&O-N \\
 22&-&-&-&0.57&O-N \\
 23&-&-&-&2.36&O-N \\
 24&-&-&-&13.22&O-N \\
 \hline
\end{tabular}
\end{table}%

The parameters for symmetry functions were adjusted by comparing the histograms of symmetry functions with different parameters. 
The overlap of the histogram was calculated for each polymorph, and eight parameters each from the point, first point-vector, and second point-vector were selected as descriptors. 
These symmetry functions were applied to trajectory of each bulk system, and the resulting calculations from each molecule at each snapshot were stored for use in the classification NN.

To train the classification NN with these sets of descriptors, 5,000 and 10,000 training samples were used  for urea and nicotinamide, respectively. 

\clearpage

\section{Graph Model Hyperparameters and Training}

The graph neural network classifier was constructed with one convolutional layer, a nodewise fully-connected layer, followed by two fully-connected layers after graph aggregation.
The graph convolution cutoff was 6~\AA.
The feature depth was 256 throughout, except for during message passing where it was bottlenecked down to 128.
Regularization was added with a dropout probability of 0.5 on all fully-connected layers, graphwise layernorm on the graph nodes, and standard layernorm on the graph embedding.
We used the Adam optimizer~\cite{kingma2014adam} with a constant learning rate of $10^{-4}$, and a batch size of 5, synthesized via gradient accumulation over 5 MD snapshots.

The train and test datasets were comprised of 1050 and 250 MD snapshots, respectively, sampled at randomly spaced time intervals, containing on average ~370 molecules each, adding up to approximately 390k total molecular environments.
Convergence studies showed similar convergence on as little as 10\% of this data, which is unsurprising, since at low temperature, most local molecular environments for a given polymorph should be very similar.

\section{Dataset Preparation}
Bulk periodic molecular dynamics trajectories of the known polymorphs of urea and nicotinamide were undertaken under the following conditions.
Simulations were undertaken using the \textsc{lammps}~\cite{LAMMPS} molecular dynamics program. 
Simulations were run for 1~ns with a time step $\Delta t = 1$~fs in the NPT ensemble using a Nos{\'e}-Hoover thermostat and barostat implemented in \textsc{lammps}~\cite{shinoda2004rapid, martyna1994constant, parrinello1981polymorphic, tuckerman2006liouville}. 

In this work, we employed the \textsc{AMBER} force field for urea and nicotinamide which relies on second generation of Generalized Amber Force Field (gaff2)~\cite{he2020fast}. 
Partial charges for urea were taken from OPLS~\cite{duffy1993urea} and for nicotinamide using RESP-charges from PBE calculations~\cite{wang2000well}.

Simulation box sizes were set as the minimum number of unit cell replicas in each cell direction to achieve at least the desired box length, where box lengths of 20~\AA\ and 40~\AA\ were used.
The 20~\AA\ samples were used in the GNN model for training on periodic bulk structures. 
The 40~\AA\ boxes were used to carve out spheres with a 30~\AA\ diameter to create molecular environments on a surface.
Surface molecules were identified as having intermolecular coordination numbers less than 20, with that value identified via coordination number histograms within several test clusters, and visually confirmed by inspection of the clusters themselves.
Initial configurations of nicotinamide gas phase clusters used  were generated in the same way and placed in large periodic boxes to simulate vacuum.

Trajectories were run at temperatures of 100~K and 200~K for urea crystal polymorphs, and 350~K for melts, and  at 100~K and 350~K for nicotinamide crystal polymorphs and 350~K  for melts. 
These temperatures were chosen to ensure that sample structures were clearly melted or crystalline for each molecule.

Melt structures were prepared starting from a stable crystal.
After relaxing the system, we gradually increased the temperature from 350~K to 2,000~K over a duration of 10 picoseconds to melt the system.
Subsequently, we reduced the temperature of the system back to 350~K on the same timescale. 
A simulation was then run for 1 nanosecond, and the resulting data was used to characterize the molten structure.

The interface structure was prepared using  form I and  IV urea structures. 
To minimize the mismatch within the system, we oriented both structures along the [001] plane, resulting in  1.7\% compression in $x$ and 1.4\% strain in $y$ of urea I and 2.8\% compression in $x$ and 0.8\% strain in $y$ of urea IV, respectively. . 
To avoid having two moving interfaces, we fixed one of the interface of form I and IV in the $z$-dimensions, then proceeded to relax the system using MD simulation.

The collected data were randomly divided into testing and training sets. 
Various sizes of training data, ranging from N = 100 to N = 50,000 unique molecular environments, were used.

\section{Symmetry Function Classifier Accuracy}

\begin{figure}
    \includegraphics[width=0.5\textwidth]{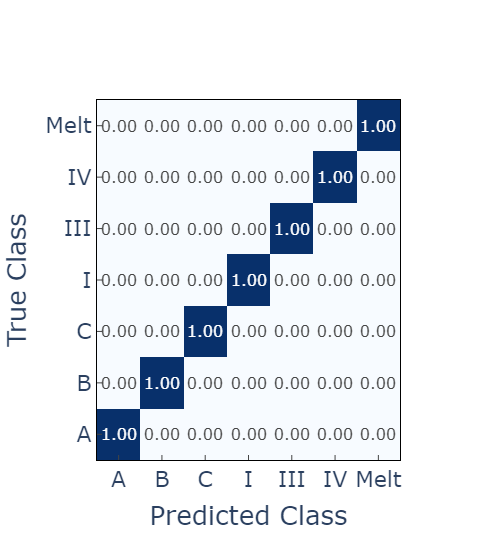}
    \caption{\label{fig:d_urea_accuracy} Confusion matrix for the symmetry function classifier on the polymorphs of urea at 200~K for crystals and 350~K for melt. Micro F1 score=1.0.}
\end{figure}

\begin{figure}
    \includegraphics[width=0.5\textwidth]{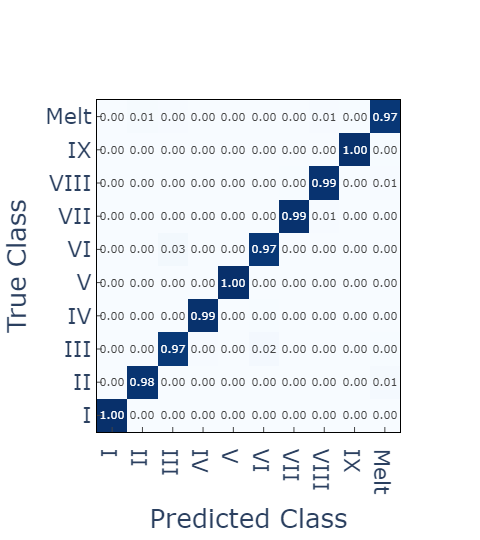}
    \caption{\label{fig:d_nic_accuracy} Confusion matrix for the symmetry function classifier on the polymorphs of nicotinamide at 350~K. Micro F1 score=0.986.}
\end{figure}

Figs.~\ref{fig:d_urea_accuracy}-\ref{fig:d_nic_accuracy} show the evaluation accuracy of the SF classifier on high temperature samples of urea and nicotinamide, respectively.
The overall accuracy is nearly perfect in both cases.
Note that the SF classifier was only trained on bulk samples, therefore surface vs. bulk classification performance is ommited in this analysis.

\newpage
\bibliography{references}